\documentclass[aps,prl,reprint,superscriptaddress]{revtex4-1}

\usepackage{graphicx}
\usepackage{epstopdf}	
\usepackage{bm}

\begin{document}


\title{Suppressed Quenching and Strong-Coupling of Purcell-Enhanced \\ Single-Molecule Emission in Plasmonic Nanocavities}


\author{Nuttawut Kongsuwan}
\author{Angela Demetriadou}
\affiliation{Blackett Laboratory, Prince Consort Road, Imperial College London, London SW7 2AZ, UK}%

\author{Rohit Chikkaraddy}
\author{Felix Benz}
\author{Vladimir A. Turek}
\author{Ulrich F. Keyser}
\author{Jeremy J. Baumberg}
\affiliation{Cavendish Laboratory, University of Cambridge, Cambridge CB3 0HE, UK}

\author{Ortwin Hess}
\affiliation{Blackett Laboratory, Prince Consort Road, Imperial College London, London SW7 2AZ, UK}%
\email[]{o.hess@imperial.ac.uk, jjb12@cam.ac.uk}



\date{\today}

\begin{abstract}
An emitter in the vicinity of a metal nanostructure is quenched by its decay through non-radiative channels, leading to the belief in a zone of inactivity for emitters placed  within $<$10nm of a plasmonic nanostructure. 
Here we demonstrate that in tightly-coupled plasmonic resonators forming nanocavities ``quenching is quenched'' due to plasmon mixing. Unlike isolated nanoparticles, plasmonic nanocavities show mode hybridization which massively enhances emitter excitation and decay via radiative channels. This creates ideal conditions for realizing single-molecule strong-coupling with plasmons, evident in dynamic Rabi-oscillations and experimentally confirmed by laterally dependent emitter placement through DNA-origami.   
\end{abstract}

\pacs{}

\maketitle

The lifetime of an excited atomic state is determined by the inherent properties of the atom and its environment, first theoretically suggested by Purcell~\cite{purcell:physrev1946} followed by experimental demonstration~\cite{goy:prl1983}. Subsequent experiments further verified this by placing atomic emitters within various optical-field-enhancing geometries~\cite{drexhage:progopt1974,kleppner:prl1981,lodahl:nature2004}. Plasmonic structures have the ability to massively enhance electromagnetic fields, and therefore dramatically alter the excitation rate of an emitter~\cite{novotny:prl2006}. However, it is well known that placing an emitter close to a plasmonic structure ($<10$nm), quenches its fluorescence~\cite{dulkeith:prl2002,kuhn:prl2006,galloway:prl2009}. Analysis by Anger~et al.~\cite{novotny:prl2006} showed this is due to the coupling of the emitter to non-radiative higher-order plasmonic modes that dissipate its energy. This `zone of inactivity' was previously believed to quench all quantum emitters. However, recent advancements have shown that an emitter's emission rate can be enhanced with plasmonic nano-antennas~\cite{farahani:prl2005,muskens:nanoletters2007,mohammadi:njp2008,kinkhabwala:naturepho2009,novotny:naturepho2011,biagioni:rpp2012,pors:acsphotonics2015,hoang:nanoletters2016}.

Generally a single emitter placed into near-contact with an optical antenna gives larger fluorescence since the antenna efficiently converts far-field radiation into a localized field and vice versa~\cite{farahani:prl2005,jun:prb2008,mohammadi:njp2008,kinkhabwala:naturepho2009}. 
This was recently demonstrated by Hoang~{\em et al.}~\cite{hoang:nanoletters2016} who showed that a quantum dot in a $12$nm nano-gap exhibits ultrafast spontaneous emission. What however remains unclear is if this enhanced emission is strong enough to allow for single emitter strong coupling.

In this Letter, we demonstrate and explain why quenching is substantially suppressed in plasmonic nanocavities, to such a degree that facilitates light-matter strong-coupling of single-molecules, even at room-temperature, as we recently demonstrated experimentally~\cite{chikkaraddy:nature2016}. This is due to: (i) the dramatic increase in the emitter excitation (similar to plasmonic antennas), and (ii) the changed nature of higher-order modes that acquire a radiative component, and therefore increase the quantum yield of the emitter. Modes in plasmonic nanocavities are not a simple superposition of modes from the isolated structures, but instead are hybrid-plasmonic states~\cite{sun:apl2010,sun:apl2011,norton:joptsocamA2008,dhawan:opticsexpress2009,vodinh:physicalchemc2010}. Hence, higher-order modes that are dark for an isolated spherical nanoparticle, radiate efficiently for tightly-coupled plasmonic structures~\cite{mertens:nanoletters2016}, significantly reducing the non-radiative decay and quenching. 
By directly comparing an isolated nanoparticle with a NPoM nanocavity (equivalent alternative nanocavities can be nanoparticle dimers with $<$3nm gap), we quantify their different radiative and non-radiative channels, explaining the mechanism that leads to suppression of quenching in plasmonic nanocavities. 
On the basis of a semi-classical Maxwell-Bloch theory of a two-level emitter, we perform Finite-Difference Time-Domain (FDTD) calculations, revealing the (spatio-temporal) emission dynamics in each system. Finally, using DNA-origami to control the position of a single emitter in the nanogap, we experimentally demonstrate the suppression of quenching in plasmonic nanocavities. 

The fluorescence rate $\gamma_{\text{em}}$ of an emitter generally depends on its excitation rate ($\gamma_{\text{exc}}$), and its radiative decay rate (i.e. quantum yield, $\eta=\gamma_{\text{rad}}/\gamma_{\text{tot}}$) as~\cite{novotny:prl2006}:
\begin{equation}
\gamma_{\text{em}}=\gamma_{\text{exc}} \, \eta = \gamma_{\text{exc}}  \left(\frac{\gamma_{\text{rad}}}{\gamma_{\text{tot}}}\right)
\end{equation}
where $\gamma_{\text{rad}}$ and $\gamma_{\text{tot}}$ are the emitter's radiative and total (Purcell factor) decay rates. 
The normalized excitation rate is governed by the field enhancement at the position of the emitter, and assuming that the environment does not affect the emitter's polarizability we have: 
\begin{equation}
\tilde{\gamma}_{\text{exc}} = \frac{\gamma_{\text{exc}}}{\gamma_{\text{exc}}^{0}}=\left|\frac{\mathbf{\hat{p}}\cdot\mathbf{E}(\mathbf{r}\text{=0})}{\mathbf{\hat{p}}\cdot\mathbf{E}_{0}(\mathbf{r}\text{=0})}\right|^2,
\end{equation}
where $\mathbf{\hat{p}}$ is the emitter's polarizability unit vector, $\mathbf{E}(\mathbf{r}\text{=0})$ is the total (incident and scattered) electric field and $\mathbf{E}_{0}(\mathbf{r}\text{=0})$ the incident field at $\mathbf{r}\text{=0}$ where the emitter is placed. The quantum yield of an emitter with radiative decay rate $\gamma_{\text{rad}}=\gamma_{\text{tot}}-\gamma_{\text{nr}}$ is then calculated assuming that non-radiative decay is due to the Ohmic losses of the metal~\cite{novotny:prl2006}: $\gamma_{\text{nr}} \propto \int_{V} \text{Re} \left\{ \mathbf{j}(\mathbf{r}) \cdot \mathbf{E}_{\text{em}}^{*}(\mathbf{r})\right\} d \mathbf{r}^3$,
where $\mathbf{j}$ is the induced current density within the volume $V$ and $\mathbf{E}_{\text{em}}$ is the field emitted by the emitter.
\begin{figure}[!h]
\centering
\includegraphics[width=\columnwidth]{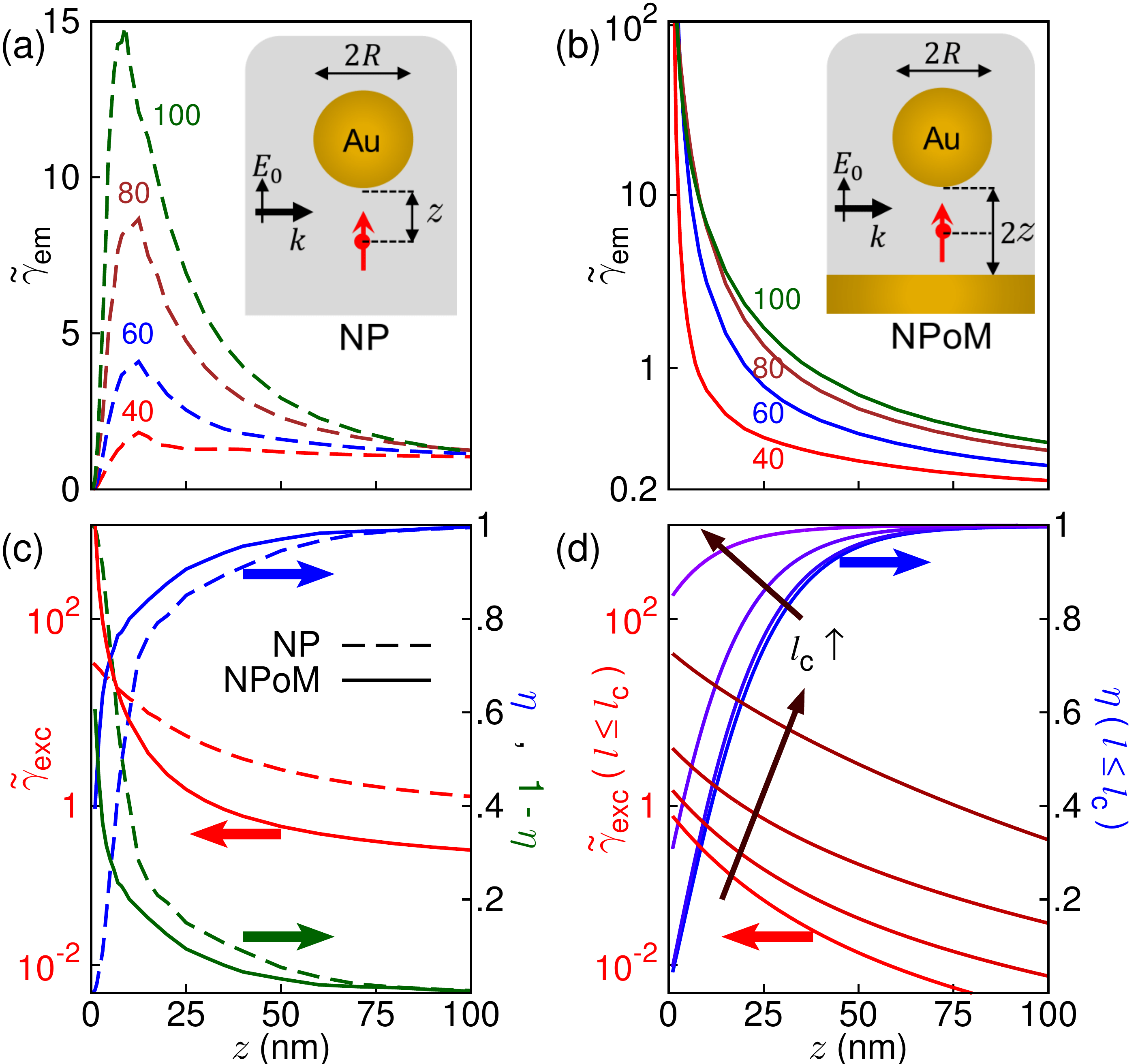}
\caption{Fluorescence rate $\tilde{\gamma}_{\text{em}}$ of an  emitter with transition $\lambda$=650nm placed at distance $z$ from (a) an isolated nanoparticle and (b) inside the NPoM nanocavity, for sphere diameters $2R$=40,60,80,100nm and with background permittivity $\epsilon_{D}$=1.96. (c) Excitation rate $\tilde{\gamma}_{\text{exc}}$ (red), quantum yield $\eta$ (blue) and $1-\eta$ (green) for an isolated nanoparticle (dotted lines) and a nanocavity (solid lines) of nanoparticle diameter $80$nm. (d) Coupling  contributions to the excitation rate (red) and quantum yield (blue) when truncating the hybridization terms at $l_{\text{c}}=2$, $3$, $5$ and $10$. \label{fig:figure1} }
\end{figure}

In the case of an isolated spherical nanoparticle (or a plasmonic nano-antenna), an emitter couples dominantly to the nanoparticle dipolar (first-order) mode. However as the emitter approaches the nanoparticle, it couples increasingly to higher-order modes, which are dark. This leads to its energy dissipation via Ohmic losses (quenching). Fig.~\ref{fig:figure1}(a) shows the normalized fluorescence rate $\tilde{\gamma}_{\text{em}}=\eta \, \tilde{\gamma}_{\text{exc}}$ for an isolated nanoparticle, calculated for a classical dipole approaching the structure, using FDTD simulations. Quenching appears when the emitter is placed at $z<10$nm, in line with previously reported results~\cite{novotny:prl2006}. By contrast, similar calculations for the NPoM nanocavity with the emitter always in the center of the nanocavity [Fig.~\ref{fig:figure1}(b)] reveal  that the emission rate increases by several orders of magnitude (note the log-scale). As $z$ decreases, the gap between nanoparticle and mirror reduces, and both plasmonic surfaces approach the emitter, but $\tilde{\gamma}_{\text{em}}$ exponentially increases.
Since the emission rate is a product of the excitation and radiative rates, we plot them separately [Fig.~\ref{fig:figure1}(c) and Fig.S1] for both an isolated nanoparticle (dashed lines) and the nanocavity (solid lines). As the emitter is progressively confined within the nanocavity, its excitation rate exponentially increases, due to the very high confinement of the plasmon modes within the gap.
Additionally, the quantum yield ($\eta$) of the nanocavity out-performs the isolated nanoparticle by more than an order of magnitude as the gap decreases. While non-local effects can affect the actual rates of emission, excitation, and quantum yield of both structures at sub-nm spacings, no significant impact is expected on their qualitative behaviour~\cite{tserkezis:nanoscale2016}. 

To illuminate the origin of these different behaviours, we adapt the analytical description of~\cite{sun:apl2010,sun:apl2011} for coupled plasmon modes. 
Isolated spherical nanoparticle follow the well-known Mie theory, but the problem of two coupled plasmonic nanoparticles is analytically more complex. It has been solved in the quasi-static limit using several techniques, such as transformation optics~\cite{li:prl2016,zhao:prl2013,luo:pnas2014} and multipole expansion~\cite{norton:joptsocamA2008,dhawan:opticsexpress2009,vodinh:physicalchemc2010,garciadeabajo:prb1999}. However, here the formalism of~\cite{sun:apl2010,sun:apl2011} is more appropriate since it  accounts for the coupling of the bare modes of the two plasmonic structures. Adapting this description for the NPoM (by approximating the mirror as a large sphere of radius $r_m$=1$\mu$m), the field enhancement in the middle of the nanocavity gap is given by~\cite{sun:apl2010}:
\begin{eqnarray}
\label{eq:field_enhancement}
& & \frac{E(r\text{=0})}{E_0} 
 \simeq  \, \alpha^{\text{NP}}\left(\frac{R}{R+z}\right)^3 \\
& &   +  \alpha^{m}\left[ 1+\sum_{l=2}^{\infty} \frac{\sqrt{\omega_{1}\omega_{l}}}{\omega_{l}-\omega-i\gamma/2} \left(\frac{l+1}{2}\right)^2  \frac{R^{(2l+1)}}{(R+z)^{(l+2)} \, r_{m}^{l-1}} \right], \nonumber
\end{eqnarray}
where $R$ is the radius of the nanoparticle, $2z$ is the gap size assuming the emitter is in the middle of the gap, and $\omega_{l}=\omega_{p}\sqrt{l/(2l+1)}$ is the resonant frequency of mode $l$, with $\omega_p$ and $\gamma$ the metal plasmon frequency and damping. The nanoparticle polarizability $\alpha^{\text{NP}}=2 \frac{(\varepsilon_{\text{Au}}-1)}{(\varepsilon_{\text{Au}}+1)}$, while the mirror polarizability $\alpha^{m}$ is given by Mie scattering (beyond the quasi-static limit) in~\cite{Kuwata:apl2003}.
The first term provides the field enhancement contribution of the nanoparticle dipole mode, the first term in square brackets is the mirror dipole mode, and the second term in square brackets is the coupling of the mirror to the higher-order modes of the nanoparticle ($l\ge 2$). In Figure~\ref{fig:figure1}(d) we plot this latter contribution of the coupling terms in equation~(\ref{eq:field_enhancement}) to the excitation rate (red lines) while truncating at increasingly higher-order modes. As the nanocavity gets smaller ($z\downarrow$), higher-order mode hybridization is needed to account for the exponential increase of the NPoM excitation rate (seen in Fig.~\ref{fig:figure1}c). Similarly, the quantum yield increases with increasingly higher-order hybridization between the two structures. 
Both these demonstrate that the mode hybridization of the coupled plasmonic structures forming the nanocavity alter the fluorescence rate of an emitter in a way that fully compensates quenching.

\begin{figure}[!h]
\centering
\includegraphics[width=\columnwidth]{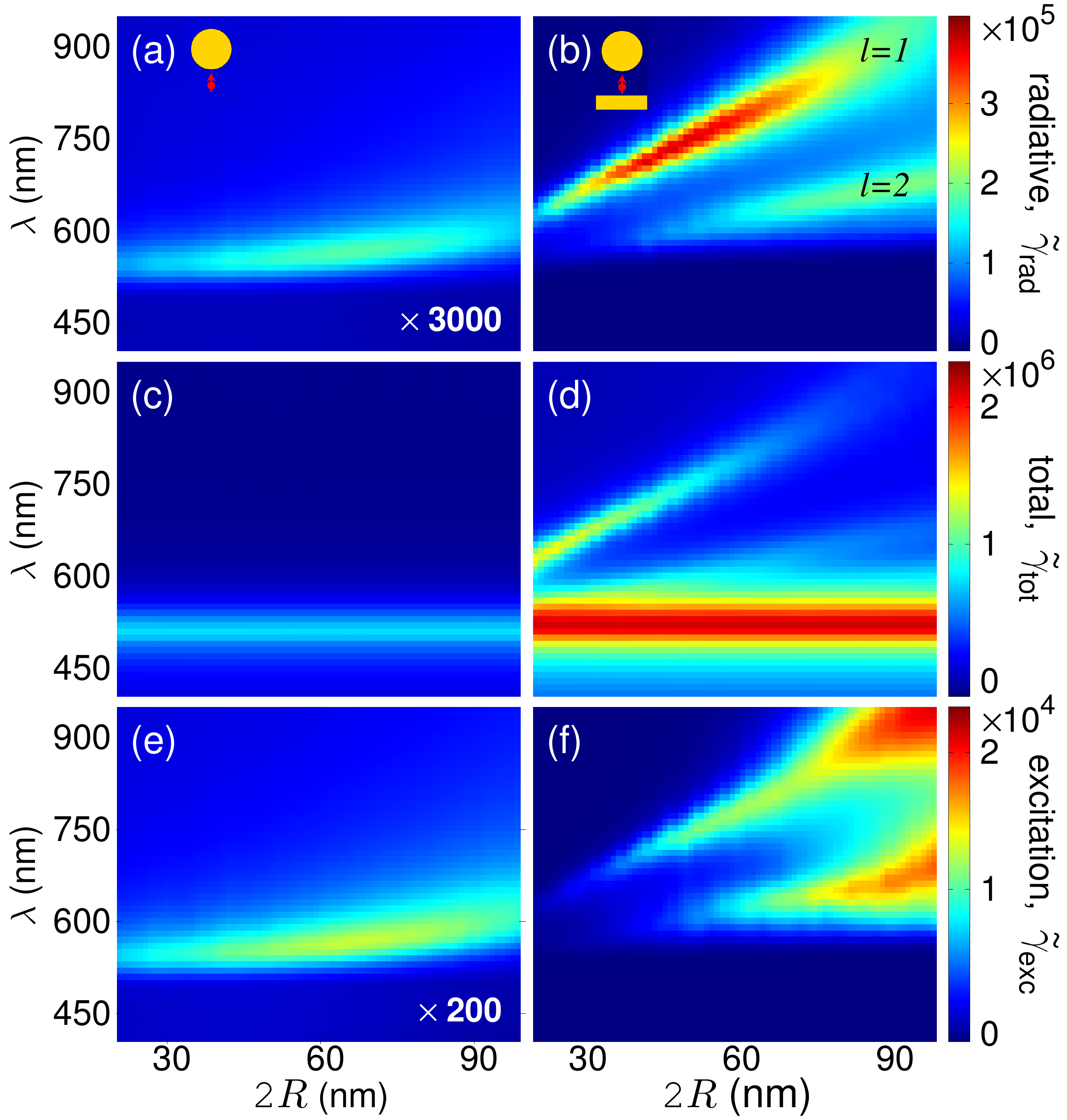}
\caption{Spectra for a vertically-oriented emitter placed  (a,c,e) $0.5$nm below an isolated nanoparticle, and (b,d,f) inside a $1$nm-wide NPoM nanocavity. (a,b) Normalized radiative decay rate $\tilde{\gamma}_{\text{r}}$, (c,d) Normalized total decay rate $\tilde{\gamma}_{\text{tot}}$ (Purcell factor), and (e,f) Normalized excitation rate $\tilde{\gamma}_{\text{exc}}$.  \label{fig:figure2}}
\end{figure}

The spectral dependence of the radiative, total, and excitation rates for both the isolated nanoparticle and the nanocavity, varying the nanoparticle diameter from $20$nm to $100$nm, show strongly contrasting behaviour (Fig.~\ref{fig:figure2}). Again the emitter is $0.5$nm from the Au surfaces, or at the centre of the $1$nm gap. Isolated quasi-static nanoparticles (with $2R<100$nm) possess diameter-independent modes [Fig.~\ref{fig:figure2}(a,c,e)]. However the resonant wavelengths of the nanocavity modes are highly dependent on the system geometry~\cite{tserkezis:pra2015,mertens:nanoletters2016} [Fig.~\ref{fig:figure2}(b,d,f)].
The NPoM radiative decay rate $\tilde{\gamma}_{\text{rad}} = \gamma_{\text{rad}}/\gamma_{0}$, normalized to the free space decay rate $\gamma_0$, is three orders of magnitude larger than for the isolated nanoparticle, with the NPoM dipole ($l=1$) mode significantly red-shifting for larger NPs. Additionally the quadrupole NPoM mode ($l=2$) strongly radiates and for larger nanoparticles has comparable radiative rates to the dipole ($l=1$) mode, in great contrast with the isolated nanoparticle.
These large $\tilde{\gamma}_{\text{rad}}$ suppress quenching, and allow strong-coupling dynamics to be radiated into the far-field.

The Purcell factor (normalized total decay rate $\tilde{\gamma}_{\text{tot}}=\gamma_{\text{tot}}/\gamma_{0}$) for both plasmonic structures shows a diameter-independent broad peak at 
$\lambda_{\text{pm}}\simeq 510$nm
 [Fig.~\ref{fig:figure2}(c,d)], which corresponds to the superposition of multiple high-order plasmonic modes, recently referred to as a `pseudo-mode'~\cite{delga:prl2014,li:prl2016}. 
However, the negligible $\tilde{\gamma}_{\text{rad}}$ at $\lambda_{\text{pm}}$ shows the large $\tilde{\gamma}_{\text{tot}}$ comes from emission coupled to the pseudo-mode decaying via non-radiative channels ($\tilde{\gamma}_{\text{tot}}=\tilde{\gamma}_{\text{rad}}+\tilde{\gamma}_{\text{nr}}$). In contrast to recent proposals~\cite{li:prl2016}, this suggests the nanocavity pseudo-mode quenches emission almost entirely via non-radiative channels, as it does for isolated nanoparticles, suppressing any way to observe possible strong-coupling dynamics. 
At the NPoM dipole and quadrupole resonant wavelengths, $\tilde{\gamma}_{\text{rad}} \sim \tilde{\gamma}_{\text{tot}} / 2$, and therefore information of the coherent energy exchange between the emitter and the plasmon modes are carried to the far-field and thus allows tracking of the hybrid states.

Additionally, the excitation rate $\tilde{\gamma}_{\text{exc}}$ of an emitter next to an isolated nanoparticle is two orders of magnitude smaller than for a $1$nm nanocavity [Fig.~\ref{fig:figure2}(e,f)]. Hence for an isolated nanoparticle where $\tilde{\gamma}_{\text{rad}} \ll \tilde{\gamma}_{\text{tot}}$, an emitter is weakly excited and heavily quenched by the pseudomode. On the other hand, the NPoM nanocavity strongly excites the emitter with the dipole/quadrupole modes, with $\tilde{\gamma}_{\text{exc}}$ increasing for larger nanoparticles, but also significant energy is  both radiated ($\tilde{\gamma}_{\text{rad}} \sim \tilde{\gamma}_{\text{tot}} / 2$) and exchanged between emitter and plasmons, allowing us to measure strong-coupling dynamics at room temperature.
This difference between the two systems is why room-temperature strong-coupling of a single emitter in plasmonic nanocavities is achievable, and has been measured experimentally~\cite{chikkaraddy:nature2016}.
\begin{figure}[!h]
\centering
\includegraphics[width=\columnwidth]{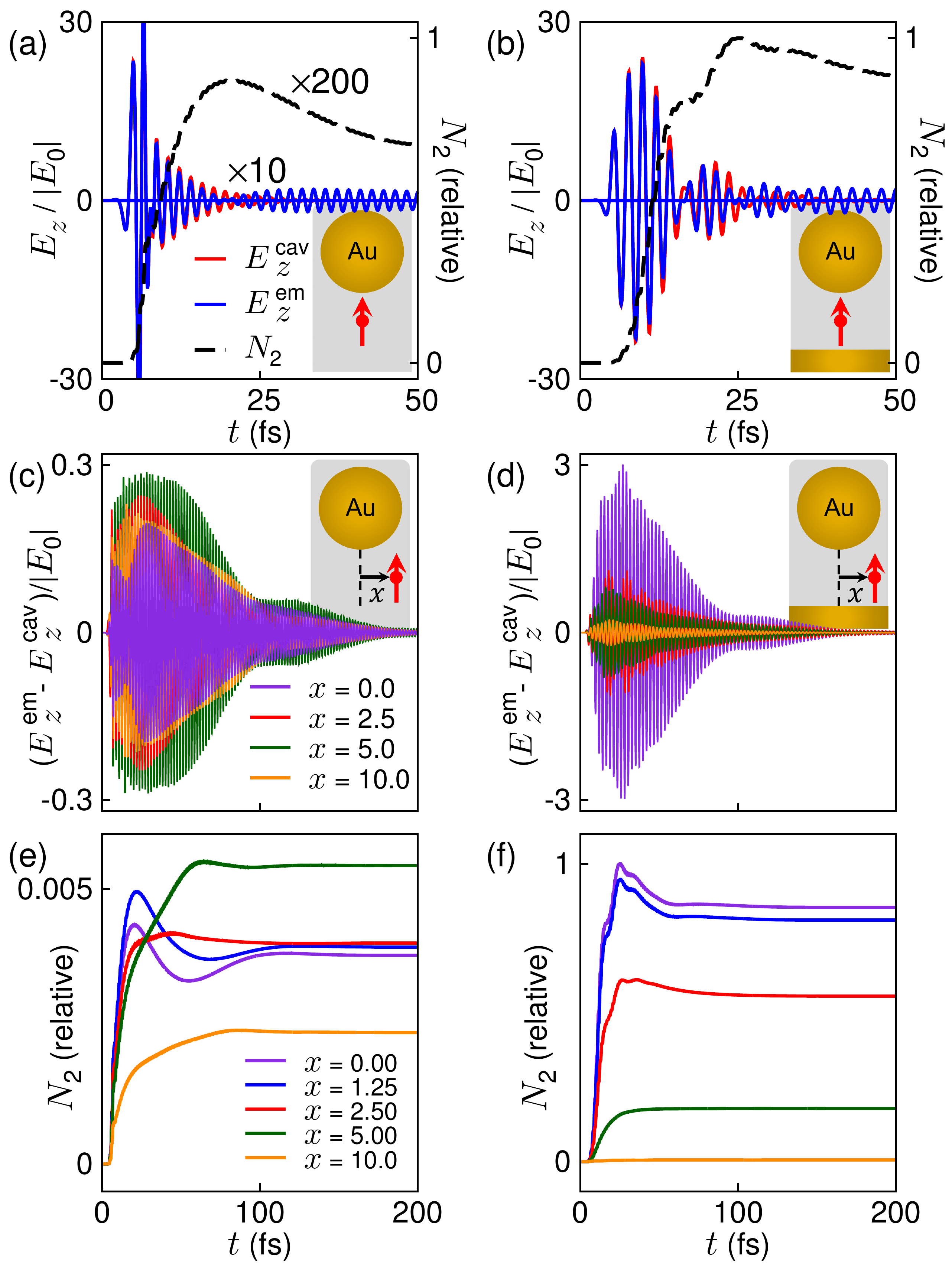}
\caption{The field $E_{z}$ and excited state population $N_{2}$ dynamics for (a) isolated nanoparticle and (b) $1$nm-wide NPoM  without (red) and with (blue) the presence of a two-level emitter $0.5$nm below the nanoparticle of diameter 40nm. The induced $E$-field from the presence of the emitter $E_{\text{em}}^{\text{ind}}=E_{z}^{\text{em}}-E_{z}^{\text{cav}}$ for (c) the isolated nanoparticle and the (d) NPoM, for an emitter placed laterally at $x$=0, 2.5, and 20nm. The population of the emitter's excited state for (e) an isoalted nanoaprticle and (f) the NPoM nano-cavity for an emitter placed laterally at $x$=0, 1.25, 2.5, 5 and 10nm. } \label{fig:figure3}
\end{figure}

While the classical calculations presented so far provide useful insight into the radiative and non-radiative decay channels of these differing plasmonic systems, they cannot reveal the spatio-temporal dynamics of an emitter coupling to the plasmons. We thus now use a dynamic two-level Maxwell-Bloch description~\cite{boyd2008} for the emitter, where the excited- ($N_{2}$) and ground- ($N_{1}$) state populations dynamics are described by:
\begin{equation}
	\frac{\partial N_2}{\partial t} = - \frac{\partial N_1}{\partial t} = -\gamma N_2 + \frac{1}{\hbar \omega_0} \left( \frac{\partial \mathbf{P}}{\partial t} + \Gamma \mathbf{P}\right)\cdot\mathbf{E}
\label{eq:rate_equation}
\end{equation}
where $\mathbf{E}$ is the local field exciting the emitter, $\mathbf{P}$ the induced polarizability, $\omega_0=2\pi/\lambda$ the transition frequency, and $\gamma$=0.66$\mu$eV and $\Gamma$=28meV are the relaxation and dephasing rates of the emitter. 
In Fig.~\ref{fig:figure3}(a), we plot the near-field $E_{z}(\mathbf{r}\text{=0})$ time evolution after a broadband pulsed excitation without (red) and with (blue) an emitter placed $0.5$nm from a nanoparticle of diameter $40$nm. We also plot the population of the excited state ($N_{2}$) on the same time-scale, which peaks at $\sim 20$fs. A qualitatively similar behaviour is observed for the NPoM [Fig.~\ref{fig:figure3}(b)] but with 4 times stronger field enhancement and 200 times larger excited state population. To clearly demonstrate the induced E-field from the emitter ($E_{\text{em}}^{\text{ind}}=E_{\text{z}}^{\text{em}}-E_{\text{z}}^{\text{cav}}$), we separate the emitter field ($E_{\text{z}}^{\text{em}}$) from the direct plasmon excitation ($E_{\text{z}}^{\text{cav}}$). In Fig.~\ref{fig:figure3}(c,d), we plot $E_{\text{em}}^{\text{ind}}$ for emitters placed at various lateral positions away from closest proximity to both the isolated nanoparticle and the NPoM. For emitters at $x<5$nm from the isolated nanoparticle, $E_{\text{em}}^{\text{ind}}$ reduces, despite the stronger field enhancement. This shows that energy from the emitter is quenched due to coupling with non-radiative higher-order modes that are confined to the vicinity of the isolated nanoparticle. For the NPoM, as the emitter approaches the nanocavity $E_{\text{em}}^{\text{ind}}$ is instead increasingly enhanced.

\begin{figure}[!h]
\centering
\includegraphics[width=\columnwidth]{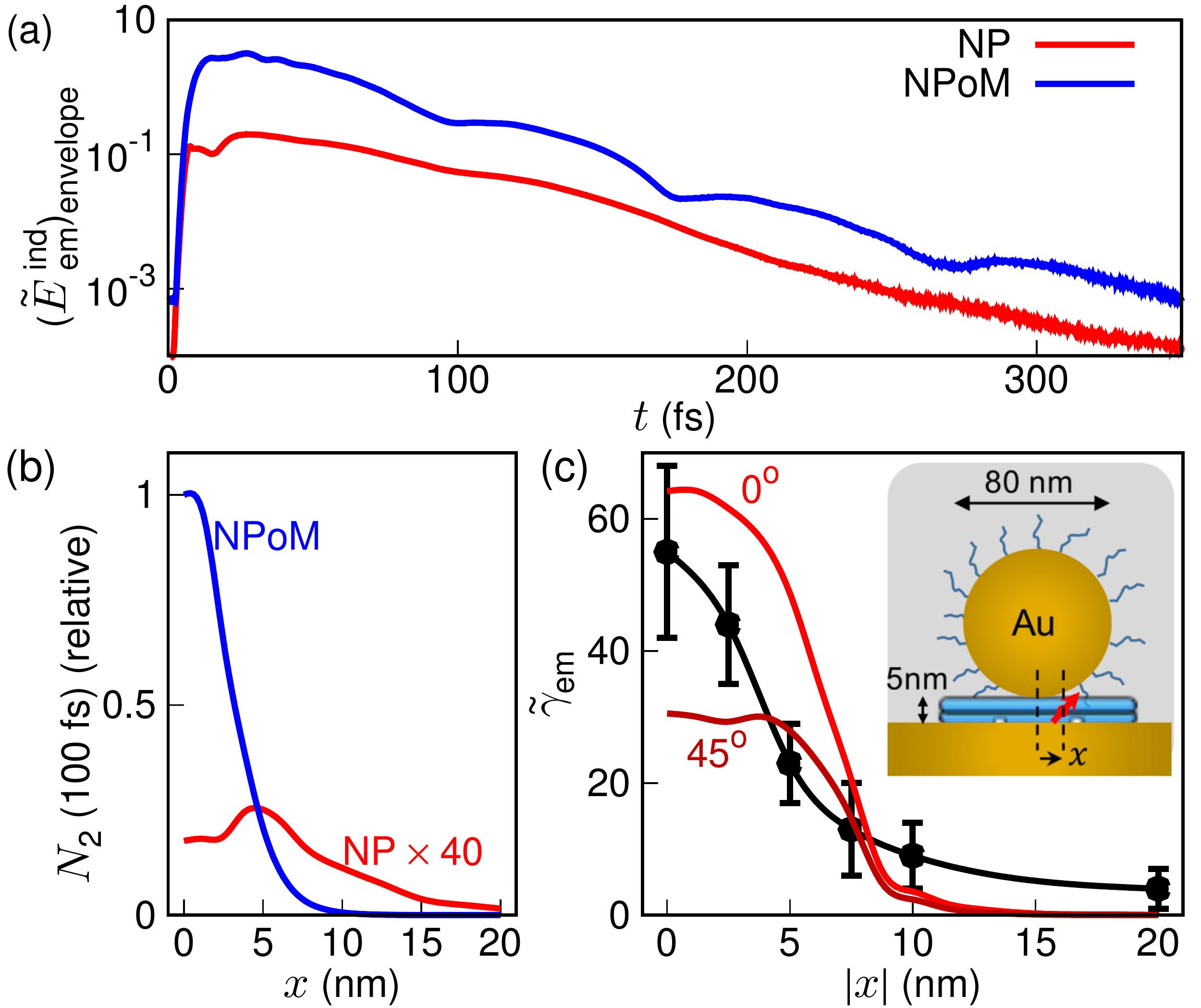}
\caption{(a) Envelope dynamics of the induced $E$-field, $\tilde{E}_{\text{em}}^{\text{ind}}=(E_{z}^{\text{em}}-E_{z}^{\text{cav}})/|E_0|$. (b) Excitation efficiency of a two-level emitter vs lateral position next to an isolated nanoparticle diameter $40$nm or $1$nm-wide NPoM. (c) Experimental (black) and numerical (red) emission intensities of a single Cy5 molecule inside a DNA-origami structure with $5$nm nanocavity gap and $80$nm diameter. Molecule is horizontally displaced by $x$ from center of nanocavity and excited by a 633nm laser. \label{fig:figure4}}
\end{figure}

Similar behaviour is observed from the excited state population dynamics [Fig.~\ref{fig:figure3}(e,f)]. For $x<2.5$nm from the lone NP, the population of the excited state is truncated by decay into the non-radiative channels, reducing it below that for an emitter at $x=5$nm, a behaviour not present for the NPoM cavity. The excited state life-times for both the isolated NP and the NPoM are shown in Fig.~S4, calculated both semi-classically and classically. This behaviour of extreme plasmonic nanocavities facilitates the strong-coupling of a single emitter at room temperature. In fact, Rabi-oscillations can be observed long after the excitation pulse is turned off for the NPoM (while almost entirely absent for the isolated NP) as clearly shown on the envelope dynamics of $E_{\text{em}}^{\text{ind}}$ at $x=0$ [Fig.~\ref{fig:figure4}(a),~S2]. The dramatic difference between the two systems is summarized in Fig.~\ref{fig:figure4}(b), where  the excited state population $N_2$ at time 100fs (after the emitter has relaxed through dephasing) is plotted for different lateral placements, $x$.

We compare our results with experiments placing a single Cy5 molecule within NPoM nanocavities formed by $80$nm diameter nanoparticles. DNA-origami~\cite{thacker:naturecom2014} is used to create a $5$nm-thick spacer and to control the emitter position at nm lateral and vertical accuracies relative to the gold-nanoparticle [Fig.~\ref{fig:figure4}(c) inset]. A 2-layer DNA-origami plate ($55 \times 60$ nm) is attached onto a gold substrate via thiol-modified staple strands. This is followed by hybridising ssDNA-functionalized gold nanoparticles with complementary overhang staple strands onto the top of the origami.\cite{thacker:naturecom2014} The position of the dye molecules with respect to the NP is varied by folding the origami with specific Cy5-modified staples at predefined positions from the centre of the NP attachment groups.  
We illuminate the nanocavity with a high numerical aperture (NA 0.8) objective, filling the back focal plane of the aperture with 633nm laser light.
 In simulations this is suitably modelled by two out-of-phase sources incident from either side of the NPoM at $55^{\circ}$ angle of incidence (inset).  The emission rates are extracted from luminescence at 690nm from $>$300 individual NPoM cavities. These intensities are referenced to a control NPoM cavity which has no dye molecule to remove any background. Note that the sub-ps emission timescales here preclude any direct measurement of emission rates, for any position of the dye molecule. 

The experimental emission rates at different lateral positions [Fig.~\ref{fig:figure4}(c), black points] quantitatively match the numerically-calculated emission rates for dipoles oriented along the $z$-axis and at $45^{\circ}$, as indicated. 
These results showing $\tilde{\gamma}_{\text{em}}(|x|)$ combine both positive and negative $x$, which are identical (Fig.~S5), placing the $x$=0 particle centre within an experimental error of $\pm2$nm.
Different DNA-origami foldings result in slightly different dipole orientations, and partial melting of the double-stranded DNA together with slight imprecision in nanoparticle placement yield the uncertainty in emitter position. 
These small variations lead to different emission intensities in different NPoMs, shown as vertical error bars in the experimental data [Fig~\ref{fig:figure4}(c)]. It is however evident that an emitter in a plasmonic nanocavity does not quench when placed in the vicinity ($<10$nm) of metal particles, but instead its emission rate enhances when moved moved towards the center of the nanocavity. 

In conclusion, we have demonstrated analytically, numerically, and experimentally that an emitter placed within a plasmonic nanocavity does not quench, despite being in very close proximity to a metal nanoparticle. This is due to (i) the enhanced excitation always present in plasmonic antennas and (ii) the acquired radiative nature of higher-order modes for extremely small gaps. The combination of the two effects both suppresses the emitter's decay into non-radiative channels and facilitates the re-emission of its energy. Plasmonic nanocavities do not quench emitters, but actually provide the necessary conditions to achieve and observe single-molecule strong-coupling with plasmons at room temperature, and many other related light-matter interactions. Extreme nanocavities at the nano-scale dimensions are fundamentally different to isolated nanoparticles and plasmonic nano-antennas with tens of nanometer gaps.

\begin{acknowledgments}
We acknowledge support from EPSRC grants EP/G060649/1 and EP/L027151/1 and European Research Council grant LINASS 320503. N.K. and A.D. contributed equally to this work.
\end{acknowledgments}

\providecommand{\noopsort}[1]{}\providecommand{\singleletter}[1]{#1}%

\end{document}